\documentclass[ prd,tightenlines,nofootinbib,superscriptaddress,twocolumn]{revtex4-1}

\usepackage{hyperref}
\usepackage{graphicx}
\usepackage[utf8]{inputenc}
\usepackage[T2A,T1]{fontenc}		
\usepackage{textcomp}
\usepackage{amsmath}
\usepackage{amssymb}
\usepackage{amsfonts}
\usepackage{mathrsfs}
\usepackage{bbm}
\usepackage{tabularx}
\usepackage{multirow}	
\usepackage{eurosym}
\usepackage{color}
\usepackage[dvipsnames]{xcolor}
\usepackage{fancyhdr}
\usepackage{fancybox}
\usepackage{stmaryrd} 
\usepackage{slashed}
\usepackage{dsfont}
\usepackage{lastpage}
\usepackage{refcount}
\usepackage{CJKutf8}

\newcommand{\me}{\mathrm{e}}
\newcommand{\mi}{\mathrm{i}}
\newcommand{\md}{\mathrm{d}}

\DeclareMathOperator{\trace}{tr}

\newcommand{\ket}[1]{{\left|#1\right\rangle}}

\renewcommand{\f}{\frac}

\def\nn{\nonumber}

\newcommand{\w}{\wedge}

\newcommand{\N}{{\mathbb N}}
\newcommand{\R}{{\mathbb R}}
\newcommand{\Z}{{\mathbb Z}}

\newcommand{\cG}{{\mathcal G}}

\newcommand{\cN}{{\mathcal N}}

\newcommand{\cV}{{\mathcal V}}

\newcommand{\cC}{{\mathcal C}}
\newcommand{\cS}{{\mathcal S}}

\newcommand{\SU}{\mathrm{SU}}

\newcommand{\be}{\begin{equation}}
\newcommand{\ee}{\end{equation}}
\newcommand{\beq}{\begin{eqnarray}}
\newcommand{\eeq}{\end{eqnarray}}
\newcommand{\bes}{\begin{eqnarray}}
\newcommand{\ees}{\end{eqnarray}}

\newcommand{\su}{{\mathfrak {su}}}

\newcommand{\la}{\langle}
\newcommand{\ra}{\rangle}

\def\eps{\epsilon}
\def\ud{{u_{\downarrow}}}
\def\uu{{u_{\uparrow}}}
\def\uud{u_{\uparrow,\downarrow}}
\def\htau{\hat{\tau}}

\begin{document}

\title{Ising Spin Network States for Loop Quantum Gravity: \\ a Toy Model for Phase Transitions}

\author{{\bf Alexandre Feller}}\email{alexandre.feller@ens-lyon.fr}
\affiliation{Laboratoire de Physique - ENS Lyon, CNRS-UMR 5672, 46 all\'ee d'Italie, 
		Lyon 69007, France}
		
\author{{\bf Etera R. Livine}}\email{etera.livine@ens-lyon.fr}
\affiliation{Laboratoire de Physique - ENS Lyon, CNRS-UMR 5672, 46 all\'ee d'Italie, 
		Lyon 69007, France}

\begin{abstract}

Non-perturbative approaches to quantum gravity
call for a deep understanding of the emergence of geometry and locality from the quantum state of the gravitational field.
Without background geometry, the notion of distance should entirely emerge from the correlations between the gravity fluctuations. 
In the context of loop quantum gravity, quantum states of geometry are defined as spin networks. These are graphs decorated with spin and intertwiners, which represent quantized excitations of areas and volumes of the space geometry.
Here, we develop the condensed matter point of view on extracting the physical and geometrical information out of spin network states: we introduce new Ising spin network states, both in 2d on a square lattice and in 3d on a hexagonal lattice, whose correlations map onto the usual Ising model in statistical physics.
We construct these states from the basic holonomy operators of loop gravity and derive a set of local Hamiltonian constraints which entirely characterize our states. We discuss their phase diagram and show how the distance can be reconstructed from the correlations in the various phases.
Finally, we propose generalizations of these Ising states, which open the perspective to study the coarse graining and dynamics of
spin network states  using well-known condensed matter techniques and results.
\end{abstract}

\maketitle

\section{Introduction} 

The search for a quantum theory of gravity remains one of the major 
challenge of theoretical physics. Even if a complete theory of quantum 
gravity is still elusive, numerous approaches
have shed lights on the problem itself and gave insights on the 
possible structure of the final theory.
In general relativity, gravity does not evolve in a pre-assumed background geometry but moulds the geometry itself: the gravitational field is encoded directly as a curved and dynamical geometry. This background independence, conjugated to the major  difficulty 
of constructing local diffeomorphism-invariant observables (see e.g.  \cite{Tambornino_2011,Rovelii_2002,Dittrich_2006}), leads to the problem of reconstructing the geometry in quantum gravity from the information stored in the quantum state of the gravitational field.

The line of research we pursue here is at the interplay between condensed matter and quantum information on one side and quantum gravity on the other: the goal is to understand how the distance can be recovered from correlation and entanglement between sub-systems of the quantum gravity state.
The issue of the emergence of distance is of course one aspect of a more general program to understand the  emergence of geometry at a semi-classical level from a purely background  independent quantum theory of gravity. To reconstruct a notion of locality and the geometry, the distance seems nevertheless to be the first natural quantity to define. 
Thus focusing on the distance, one could extract it from the 2-point correlation function assuming that the inverse square law holds. This perspective is quite similar to the point of view of spectral geometry and non-commutative geometry, in which distances, and more generally the geometry, emerge from the spectral analysis of the 2-points function, 
given respectively by the Laplacian and Dirac operators \cite{Kempf_2008,Connes_2007}.
Like in condensed matter physics, we expect the behavior of the correlations to
depend on the considered phase. While the typical behavior of correlations 
is an exponential decay, the interesting regime for gravity, namely an algebraic 
decay, would be attained in critical regimes or algebraic phases of some specific 
model like the $XY$ model.
In this paper, we would like to propose a simple framework to test these ideas in the context of loop quantum gravity and we will introduce toy quantum states, which we dub Ising spin network states, whose correlations and phase diagram can be mapped onto the standard Ising model in statistical physics. This will allow us to discuss the relation between correlation and geometry in a well-under-control environment.

\medskip

Loop quantum gravity is  one of the most developed approach to quantum gravity and proposes
a non-pertubative canonical quantization of general relativity (for
textbooks, see \cite{Rovelli_book,Rovelli_Vidotto_book,Thiemann_book}).
The physical picture that emerges from this theory describes 
quantum space-time as a discrete geometry where areas and volumes 
operators have a discrete spectrum. The formalism is based on a $3+1$ 
formulation of general relativity in terms of the Ashtekar-Barbero  
connection and a densitized triad, thus describing gravity as $SU(2)$ gauge 
field theory, in fact a topological constrained field theory. In the quantum theory, spin network states form a basis of the 
wave functions that describes the quantum state of $3d$ space at the kinematical level. Physical spin network states will be the solution of the Hamiltonian constraints, which implement the Einstein equations and generate the dynamics. 

The core of our investigation will be on correlations and entanglement entropy
on spin network states.
In the long run,  the aim is to analyze the typical  -possibly universal- structure of spin network states with well-behaved 
algebraic correlations and satisfying an area law entropy.
Correlations and especially entropy are of special importance for the 
understanding of black holes dynamics. Indeed, understanding the 
microscopic origin of black holes entropy is one the major test of 
any attempt to quantify gravity and entanglement between the horizon
and its environment degrees of freedom appears crucial (for recent 
developments in loop quantum gravity, see \cite{Perez_2014,
Bianchi_2012}).
The usual spin network basis states are eigenstates of the area and volume operators, but they carry trivial correlations 
because of their factorized structure with no entanglement at all between 
intertwiners degrees of freedom. We therefore look for more involved spin network states, correlating the spin and intertwiner degrees of freedom living on the whole graph.
Having in mind statistical and condensed matter physics where studies 
of correlations are fairly advanced both numerically and analytically 
\cite{Pelissetto_2002}, the method we propose is to construct pure 
spin network states from well-known condensed matter models and 
possibly highlights some relevant lessons for loop quantum gravity.
A similar endeavor is currently being investigated by Bianchi  and collaborators, similarly using condensed matter methods, who introduce squeezed spin network states by a Bogoliubov transformation and focus on the evaluation of their entanglement  entropy \cite{Bianchi_2015}.

Ideally, testing the relation between correlations and distance requires
physical states solutions of the Hamiltonian constraints of loop quantum gravity. However, these constraints are rather complicated and, despite several proposals, no consensus has been reached on the structure of physical states solving those constraints at the quantum level (see e.g. \cite{Bonzom_2011} for a review of the recent research efforts on the quantum dynamics of loop gravity). To circumvent this issue, we propose a different strategy: build test spin network states on regular lattices, with an {\it a priori} notion of distance, and compare that natural lattice distance with the reconstructed distance emerging from the correlation carried by the quantum states. Working in such a controlled setting should allow us to thoroughly illustrate the link between distance and correlation, test the viability of the proposal of reconstructing the distance entirely from correlations and understand the structures of well-behaved states supporting this proposal.

\medskip

A spin network state is defined on a graph, dressed with spins on the edge and intertwiners at the vertices. A spin on an edge $e$ is a half-integer $j_{e}\in\N/2$ giving an irreducible representation of $\SU(2)$ while an intertwiner at a vertex $v$ is an invariant tensor, or singlet state, between the representations living on the edges attached to that vertex. Spins and intertwiners  respectively carry the basic quanta of area and volume.
We build our spin network states based on three clear simplifications:
\begin{enumerate}

\item We work on a fixed graph, discarding graph superposition and graph changing dynamics for now, and we will focus on working with a fixed regular lattice.

\item We freeze all the spins on all the graph edges. We fix them to their smallest possible value, $\f12$, which correspond to the most basic excitation of geometry in loop quantum gravity, thus representing a quantum geometry directly at the Planck scale.

\item We restrict ourselves to 4-valent vertices, which represent the basic quanta of volume in loop quantum gravity, dual to quantum tetrahedra.

\end{enumerate}

Since the spins are frozen, the only degrees of freedom left are the intertwiners living at the 4-valent vertices. At each vertex, these recouple between 4 spins $\f12$, which corresponds to a two-dimensional Hilbert space or two-level quantum system with pairs of spins recoupling to a spin 0 or 1. This is completely equivalent to working with a qubit at each vertex.
Qubits,  as the elementary physical systems encoding information, are the basic 
ingredients of quantum information and quantum computing and are thus the basic building blocks of condensed matter models.
Therefore these simplifications provide us with the perfect setting to map spin network states, describing the Planck scale quantum geometry, to qubit-based condensed matter models. Such models have been extensively studied in statistical physics and much is known on their phase diagrams and correlation functions, and we hope to be able to import these results to the context of loop quantum gravity.
One of the most useful model
is the Ising model whose relevance goes from modeling binary mixture 
to the magnetism of matter. We thus naturally propose to construct and 
investigate Ising spin network states.

\medskip

In three space dimensions, the natural 4-valent regular lattice is the diamond lattice, which is a honeycomb lattice and can be seen as dual to a tetrahedral discretization of space. It is different from the usual cubic lattice used in lattice gauge theory,  discretized gravity (\`a la Regge) or numerical general relativity, but we believe it is more natural and better suited to the loop quantum gravity context. This is nevertheless a brand new proposal to construct spin networks on the regular 3d diamond lattice. We will define $3d$ Ising spin network states on that lattice and analyze their properties.
For mathematical simplicity and representation purposes, we will first focus 
on their two-dimensional counterpart and define $2d$ Ising spin network states  on a square lattice. In fact, 
many obtained results are straightforwardly generalized to $3d$. The 
$2d$ Ising model on a regular square lattice without magnetic field has 
been solved exactly \cite{Baxter_book} and exhibits a phase transition with
algebraically decaying correlations at the critical coupling. The phase diagram can be studied using 
quantum field theory methods. Since 
away from the transition the correlations decay exponentially, the 
relevant scenario for gravity is indeed at the critical coupling which 
also has the major advantage to possess a continuum limit.

Thus this state appears to fall into the category of the test states we are
looking for. Moreover, this method of investigation allows us to understand
the typical structure of such states  in terms of 
the holonomy operators of loop quantum gravity: non-trivial correlations seem to require holonomies around arbitrarily extended loops. We also show that this
state is a unique solution of a set of local Hamiltonian
constraints illustrating how local constraints can give rise to extended 
holonomies and to long range correlations, a common feature of condensed 
matter systems.
 
The paper is structured as follows. Section \ref{Def}  reviews the definition of spin network and analyze the structure of 4-valent intertwiners between spins $\f12$ leading  to the effective two-state systems
which we use to define the Ising spin network states.
%
Different equivalent 
definitions are given in terms of the high and low temperature expansions 
of the Ising model. The loop representation of the spin network is then
obtained and studied as well as the associated density which gives 
information about parallel transport in the classical limit. Section \ref{hamil}
introduces a set of local Hamiltonian constraints for which the Ising state
is a unique solution and elaborates on their usefulness for understanding the 
coarse-graining of spin network \cite{Livine_2014} and the dynamic of 
loop quantum gravity. Section \ref{phase_diag} discusses the phase diagram 
of our Ising states and their continuum limit as well as the distance 
from correlation point of view. Section \ref{Conc} end 
this paper with discussion and perspective.

\section{Ising spin network state}
\label{Def}

\subsection{Definition}
\label{def}

\begin{figure}[h]
  \centering
  \includegraphics{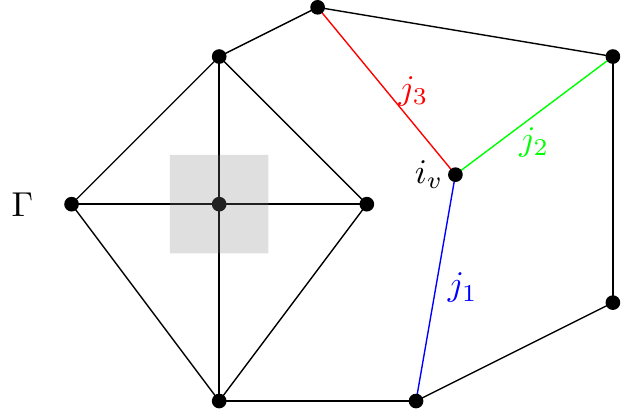}
  \caption{Example of a colored graph $\Gamma$ with spins $j_{1,2,3}$ at the source node $v$ where the interwiner $i_v$ belongs. The shaded 
		region represent an example of the dual geometry encoded in the spin 
		network state.}
  \label{spin_network}
\end{figure}

In loop quantum gravity, kinematical states are cylindrical functionals of the Ashtekar-Barbero $\SU(2)$ connection that depend only on its 
holonomies along a set of paths which define an oriented graph
$\Gamma$.  We call $E$ the number of edges of te graph $\Gamma$ and $V$ its number of vertices. 
These wave-functions of the geometry  depend on $E$ group elements in $\SU(2)$. We further require those states to be gauge invariant, which means an invariance under  $\SU(2)$ transformations at each vertex. The Hilbert space of wave-functions
\footnote{In fact, the true states of geometry in loop quantum gravity are defined as a sum over all possible graphs $\Gamma$ thanks to a  projective limit\cite{Ashtekar:1994mh} of  graphs $H_{\text{kin}} = \lim\limits_{\Gamma \rightarrow +\infty} 
H_{\Gamma}$. The measure on this space of generalized connections 
$\overline{A}$ is given by the Ashtekar-Lewandowski measure $\md \mu_{AL}$. 
So $H_{\text{kin}}\simeq L^2(\overline{A}, \md \mu_{AL})$. This projective limit allows to make sense rigorously of superposition of states living on different graphs.}
on the graph $\Gamma$ is then $L^2(SU(2)^E/SU(2)^V, \md \mu)$ where the scalar product is defined by the $\SU(2)$ Haar measure $\md \mu$.
A basis of this Hilbert space is given 
by spin networks states using the Peter-Weyl theorem stating that a basis of $L^{2}(\SU(2),\md\mu)$ is given by the Wigner matrices of the $\SU(2)$ group element in all possible irreducible unitary representations. On each link 
$e$ of the graph lives a spin $j_e$ associated to the area degrees of 
freedom. At each vertex $v$ lives an $SU(2)$ invariant 
tensor called intertwiner $i_v \in \text{Inv}_{SU(2)} \left( \otimes_{e\ni v} 
\cV^{j_e} \right)$, with $\cV^{j_e}$ is the $(2j_{e}+1)$-dimensional Hilbert space for the spin $j_e$ representation.
These spin network basis states, as illustrated on fig.\ref{spin_network}, define the basic excitations of the quantum geometry and they are provided with a natural interpretation in terms of discrete geometry with the spins giving the quanta of area and the intertwiners giving the quanta of volume. This interpretation is most clear in the twisted geometry picture \cite{Freidel:2010aq}.

More explicitly, the spin network state $\ket{\Gamma, j_e, i_v}$, based on the graph $\Gamma$ and dressed with spins $j_{e}$ and intertwiners $i_{v}$, defines a  wave function $\psi(g_e)$ on the space of discrete connections of $SU(2)^E/SU(2)^V$
\begin{align}
\psi^{\{j_{e},i_{v}\}}(g_e) = \trace{\bigotimes_e D^{j_e}(g_e) \otimes \bigotimes_v i_v} 
\end{align}
where the trace contracts the intertwiners with  the Wigner matrix $D^{j_e}(g_e) $ representing 
the group element $g_e$ along the links between vertices. The 
scalar product between those states are obtained using the orthogonality
relation of the Wigner matrices:
\begin{align}
\langle \Gamma, \overline{j}_e, \overline{i}_v |
\Gamma, j_e, i_v \rangle =
\prod_e \frac{\delta_{j_e,\overline{j}_e}}{2j_e + 1}
\prod_v \langle j_v | i_v \rangle
\end{align}
Those spin network states, being tensor product of intertwiners, have
trivial correlations.
We would like to build quantum superposition of those basis spin network states that would cary non-trivial correlation. What we will develop below is a method to construct such states whose correlation structure maps onto the Ising model. We will then be able to translate the known quantities and results of statistical physics to the context of gravity. 

\medskip

As outlined in the introduction, we propose to define Ising spin network states by freezing the spins to their smallest value, $j_{e}=1/2$, that is the smallest quanta of area at the Planck scale, and to focus on 4-valent intertwiners. These 4-valent vertices will be organized along a regular lattice. We will consider the 3d diamond lattice and the 2d square lattice. We will focus on the square lattice as first step of analyzing the 3d case. Since most of  the results  obtained in 2d are straightforwardly generalized to 3d as we will see, we postpone the discussion of the 3d Ising spin network to the later section \ref{3d_case}.
Those regular lattices carry a natural \emph{a priori} 
notion of distance but nothing says it corresponds 
to a physical distance. We will strive to compare this lattice distance with the emergent distance reconstructed from the physical correlation carried by the quantum state.
\begin{figure}[h]
  \centering
  \includegraphics{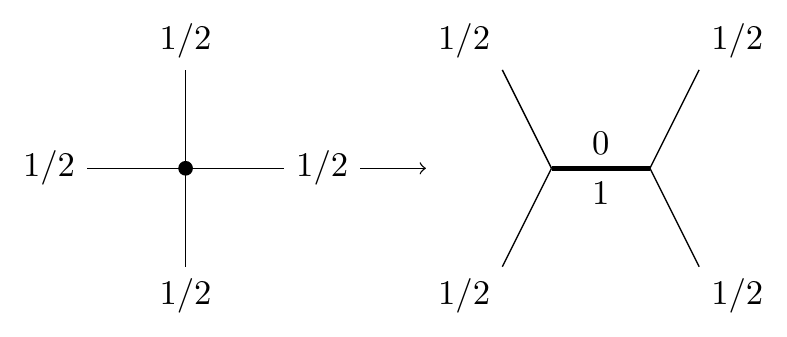}
  \caption{Graphic representation of the 4-valent spin $1/2$ and one
	      of its possible decomposition into a spin $0$ and $1$ states.}
  \label{intert}
\end{figure}

\begin{figure}[h]
  \centering
  \includegraphics{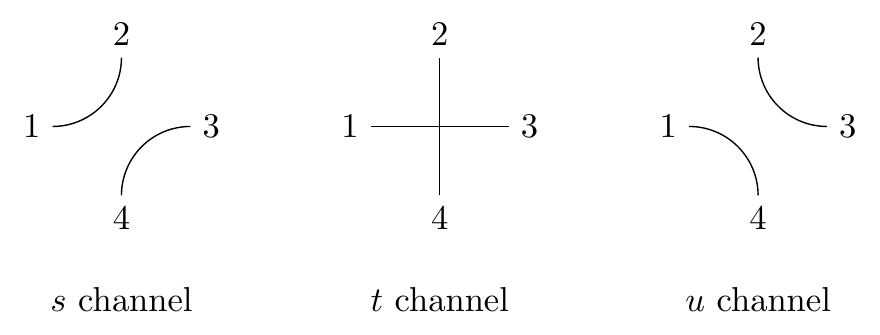}
  \caption{Graphic representation of the different decompositions of 
  		the 4-valent spin $1/2$ intertwiner. A spin $0$ or $1$ state
  		can be associated to each of these decompositions.}
  \label{mand_intert}
\end{figure}

In this setting, the space of 4-valent interwiners between four spins $1/2$ is two dimensional: 
it can be decomposed into  spin $0$ and spin $1$ states by combining the spins by pairs, as on figure 
\ref{intert}.
Different  such decompositions exist and figure \ref{mand_intert} gives a graphical 
representation of them. There are three such decompositions, depending on which spins are paired together, which we dub $s$, $t$ and $u$ channels as in particle physics and quantum field theory.
Writing $\cV$ for the 2-dimensional Hilbert space of a spin $1/2$, we denote its basis states  $\left( \ket{\uparrow}, \ket{\downarrow} \right)$.
We call $\ket{0_{s}}$ and $\ket{1_{s}}$ the spin $0$ and $1$  states in the $s$ channel. They can be explicitly written in terms of the up and down states of the four spins:
\begin{align}
\label{intert_state_12}
\ket{0_{s}} &= \frac{1}{2}
	\left( 
		\ket{\uparrow \downarrow \uparrow \downarrow} +
		\ket{\downarrow \uparrow \downarrow \uparrow} - 	
		\ket{\uparrow \downarrow \downarrow \uparrow} -
		\ket{\downarrow \uparrow \uparrow \downarrow}
	\right)	 \nonumber \\
\ket{1_{s}} &= \frac{1}{\sqrt{3}}
		\left(
		\ket{\uparrow \uparrow \downarrow \downarrow } + 
		\ket{\downarrow \downarrow \uparrow \uparrow }
		\right. \nonumber \\
		&-
		\left.
		\frac{
			\ket{\uparrow \downarrow \uparrow \downarrow} + 
			\ket{\downarrow \uparrow \downarrow \uparrow} +
			\ket{\uparrow \downarrow \downarrow  \uparrow} +
			\ket{ \downarrow \uparrow\uparrow \downarrow} 
		}{2}
		\right)
\end{align}
Those two states obviously  form a basis of the intertwiner space. We give can the transformation matrices between this basis and the two other channels:
\begin{align}
\label{chgt_basis}
\begin{pmatrix}
\ket{0_t} \\ \ket{1_t}
\end{pmatrix} &= 
\frac{1}{2}
\begin{pmatrix}
1 & \sqrt{3} \\
 \sqrt{3} & -1
\end{pmatrix}
\begin{pmatrix}
\ket{0_s} \\ \ket{1_s}
\end{pmatrix}
 \nonumber \\
\begin{pmatrix}
\ket{0_u} \\ \ket{1_u}
\end{pmatrix} &= 
\frac{1}{2}
\begin{pmatrix}
-1 & \sqrt{3} \\
- \sqrt{3} & -1
\end{pmatrix}
\begin{pmatrix}
\ket{0_s} \\ \ket{1_s}
\end{pmatrix}
\end{align} 
The intertwiner basis that actually interests us is not attached to one of these channels but is defined in terms of the square volume operator $\hat{U}_v$ of loop quantum gravity. Since the spins, and thus area quanta, are fixed, the only freedom left in the spin network states are the volume quanta defined by the intertwiners. This will provide the geometrical interpretation of our spin network states as excitations of volumes located at each lattice node. For a 4-valent vertex, this operator is defined as:
\begin{align}
\hat{U}_v = \left( \frac{\sqrt{2}}{3} \right)^2 
\left( 8\pi G  \hbar \gamma \right)^3
\left( -\mi[\vec{J_1}.\vec{J_2}, \vec{J_1}.\vec{J_3}] \right)
\end{align}
where $\vec{J_i}$ are the spin operators acting on the $i$ link. For clarity, 
we will for the remaining of this paper use the natural quantum gravity units for which 
$G = c = \hbar = 1$.
This operator is Hermitian but not positive; it also registers the space orientation (or more precisely the orientation of the bivector space over $\R^{3}$, which is itself isomorphic to $\R^{3}$). The volume itself can then obtained by taking the squareroot of the absolute value of $\hat{U}$.
Geometrically, 4-valent intertwiners are interpreted as representing quantum tetrahedron, which become the building block of the quantum geometry in loop quantum gravity and spinfoam models \cite{Baez_1999,Barbieri:1997ks}.
Using the change of basis relations  \eqref{chgt_basis}, $U_v$ takes the following form in the $s$ channel basis:
\begin{align}
\hat{U}_v= \mi \frac{\sqrt{3}}{4}
\begin{pmatrix}
0 & 1 \\
 -1 & 0
\end{pmatrix}
\end{align}	
The eigenstates $\ket{u_{\uparrow,\downarrow}}$ can be obtained directly has 
$\ket{\uud} 
= \frac{1}{\sqrt{2}} 
	\left( 
		\ket{0_{s}} \mp i \ket{1_{s}}
	 \right)$
with associated eigenvalues in Planck units $\pm\frac{\sqrt{3}}{4}$ . 
This (square) volume is the smallest non trivial possible value of a 
chunk of space. The only freedom is the orientation of the chunk of volume corresponding to the intertwiner state.

\medskip

We consider these two oriented volume states $\ket{\uud}$ as the two levels of an effective qubit. 
Let us point out that these two level states are a priori only defined up to phases, and we could actually choose arbitrary phases $\ket{\uud} \,\rightarrow \,\me^{i\theta_{\uparrow,\downarrow}}\,\ket{\uud}$. Choosing $\sqrt{2}\ket{\uud} =\ket{0_{s}} \mp i \ket{1_{s}}$ fixes those phases and actually selects the $s$-channel over the $t$ or $u$ channels.

We can now define a pure   spin network state which maps its quantum 
fluctuations on the thermal fluctuations of a given classical statistical
 model -for instance the Ising model-  by 
\begin{align}
\label{def_equa}
\ket{\psi} &= \sum_{\{\sigma_v=\pm\}} 
			A[\sigma_v] 
			\me ^{\mi \Theta [\sigma_v] }
			\psi^{\sigma_v 1/2} \nonumber \\
\textrm{with}\quad
A[\sigma_v] &= \me^{\frac{J}{2} \sum_{\langle v,v' \rangle} \sigma_v \sigma_{v'}
		 + \f 12 \sum_v {B_{v}}\sigma_v }
\end{align}
where we sum over all the possible spin $\sigma_v$ configurations modulated
by an arbitrary phase $\Theta[\sigma_v]$  and an amplitude $A[\sigma_v]$ which has been chosen to be an Ising
nearest-neighbors model with a coupling constant $J$ and magnetic
fields $B_{v}$. We have mapped the Ising spin $\sigma_v=\pm$ onto the space orientation of the square volume eigenvalues $|\uud\ra$.
The state $\psi^{\sigma_v 1/2}$
represents a particular configuration of the spin network and the full
state is a quantum superposition of them all. Defined as such,  the state is unnormalized but
its norm is easily computed to be the Ising partition  function $Z_{\text{Ising}}$:
\be
Z_{\text{Ising}}
=
 \sum_{\{\sigma_{v}\}} \me^{J\,\sum_{e}\sigma_{s(e)}\sigma_{t(e)}}
=
 \sum_{\{\sigma_{v}\}} \me^{J\, \sum_{\langle v,v' \rangle} \sigma_v \sigma_{v'}}
\ee
using the usual condensed matter notation $\langle v,v' \rangle$ for nearest neighbor vertices or the usual loop gravity notation $s(e),t(e)$ respectively for the source and target vertices of every (oriented) edge $e$ of the graph.

In principle we could define such states with any 
amplitude involving Ising spins and generalize our Ising spin network states to any other condensed matter models built from 2-level systems.

The intertwiner states living at each vertex are now entangled and carry non-trivial correlations.
More precisely this state exhibits Ising correlations between two vertex $i$, $j$:
\begin{align}
\langle \sigma_ i \sigma_j \rangle  = 
\frac{1}{Z_{\text{Ising}}} \sum_{[\sigma_v]} 
	\sigma_i \sigma_j |A[\sigma_v]|^2
\end{align}
Those correlations are between two volume operators at different vertex 
which are in fact components of the 2-point function of the gravitational field. So 
understanding how those correlations can behave in a non-trivial way 
is a first step toward understanding the behavior of the full 2-point gravity correlations and for instance recover the inverse square law of the propagator.

We see that in order to emulate correlations of a classical statistical
system, the amplitudes appearing in the pure state must be the square 
root of the classical Boltzmann factor.
More generally, a spin 
dependent phase $\Theta[\sigma_v] $ could be introduced
in the definition of the state itself as done in Eq.\ref{def_equa}.
This phase could for example consist in complex valued local magnetic fields, which occur  in the Lee-Yang zeroes theorem and are relevant to some decoherence models \cite{PhysRevLett.109.185701}.
For instance, if we change the phases in the definition of the two states $\ket{\uud}$, this would clearly change the Ising spin network state defined above, but the phases would entirely be re-absorbed into the Ising Hamiltonian as purely imaginary magnetic fields. Such a 
phase term doesn't actually change the spin correlations, 
$\langle  \cdot \rangle_{{\Theta}} = 
\langle \cdot \rangle_{{\Theta = 0}}$.
They will nevertheless affect the expectation values of dual operators that shift the spins 
and will affect the Hamiltonian constraints satisfied by the state as we will see below. 
 For now,
we consider this phase to be equal to unity for the sake of simplicity.

Similar states have been studied in a quantum information approach
to condensed matter physics and appear to have nice properties with 
respect to entanglement entropy or even for quantum computation 
purposes \cite{Cirac_2006,Cirac_2008}. The spirit of this approach 
consists in constructing quantum states with well controlled physical 
properties and then understand them as a ground state of a particular
dynamic. This is the same perspective we are advocating for loop 
quantum gravity here.

\subsection{Low and high temperature expansion definition}
\label{ht_lt}

In order to better understand the structure of the Ising spin networks and how to build them from the basic loop quantum gravity operators, we look at them for the perspective of the low and high temperature loop expansions of the Ising model.
This is also interesting from the point of view that the duality between the loop expansions at low and high temperatures allows to characterize exactly the critical point of the Ising model, especially for the 2d square lattice, which is self-dual (see e.g. \cite{Baxter_book}).

\smallskip

\paragraph{Low temperature expansion}

\begin{figure}[h]
  \centering
  \includegraphics{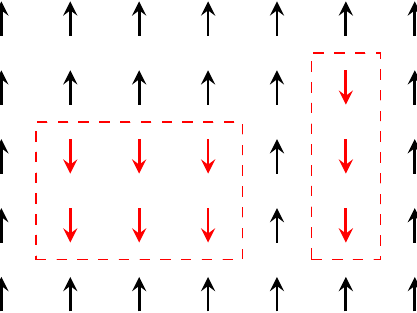}
  \caption{An example of a cluster configuration $C\subset\Gamma$ composed of two groups of Ising down spins $\ud$ amidst the up spins $\uu$. The cluster  boundary is made of two loops on the dual lattice, which form an even subgraph.}
  \label{low_T}
\end{figure}

The partition function of the Ising model admits different representations
which are more relevant in different temperature regimes. We restrict
ourselves here to the case with no local magnetic field. In the low
temperature regime, typically below the critical temperature, the ground state of the system is an ordered phase. It is then 
natural to extract the ground state contribution from the partition 
function and focus explicitly on the excitation contributions. Such a 
representation is called the low temperature expansion, or cluster expansion. For a planar graph, as the square lattice, it reads 

\bes
Z_{\text{Ising}}
&=&
 \sum_{\{\sigma_{v}=\pm\}} \me^{J\,\sum_{e}\sigma_{s(e)}\sigma_{t(e)}}
\\
&=&
2 \me^{\frac{zJ N}{2}} \sum_{C\in\cC} \me^{- 2 J P_C}
\,=\,
2 \me^{\frac{zJ N}{2}} \sum_{\gamma^{*}\in\cG^{*}} \me^{- 2 J P_{\gamma^{*}}}\nn
\ees
where $z$ is the valence or order of the graph's nodes, fixed here at $z=4$ for the square lattice.
We are summing over all clusters $C$, that is all subsets of vertices of the graph $\Gamma$. The expansion is obtained by distinguishing the up spins from the down spins. A cluster $C$ is equivalent to a even subgraph $\gamma^{*}\subset\Gamma^{*}$ of the dual lattice, that  is such as the valence of each node of the subgraph is even, i.e. 0 (the node does not belong to the subgraph), 2 (the subgraph goes through that node) or 4 (which is the maximal value). This even subgraph on the dual graph is simply the boundary of the cluster, as illustrated on figure  \ref{low_T}. $P_{C}=P_{\gamma^{*}}$ is the number of edges of the dual subgraph or equivalently the total perimeter of the cluster.
One can decompose an even subgraph in terms possibly intersecting loops, with loop intersection corresponding to the 4-valent nodes of the subgraph.

This low temperature expansion leads to  an alternative definition of the Ising
spin network, which we note $\ket{\psi_{\text{LT}}}$. We  start from the totally ordered 
state $\ket{\uu\uu \cdots}$ and act with an
operator that flip all spins inside a given cluster  $C$. This reads:

\begin{align}
\ket{\psi_{\text{LT}}} &= \sqrt{2} \me^{J N} 
		\sum_{C} \me^{- J P_C} \ket{C} \nonumber \\
\ket{C} &\equiv \hat{\tau}^C_x \ket{\uu\uu \cdots} 
	      \equiv \prod_{v\in C}  \hat{\tau}^v_x 
	      \ket{\uu\uu \cdots} 
\end{align}
where $\hat{\tau}$ is the notation used for the Pauli matrices.
$\hat{\tau}_{x}$ is the flip operator that sends the up spin $\ket{\uu}$ on a down spin $\ket{\ud}$ and vice versa. It can be constructed in terms 
of geometric  operators, from area and scalar product operators acting on the intertwiner space, as shown in appendix \ref{sum_states}.


\smallskip

\paragraph{High temperature expansion}

In the high temperature regime, their exist various expansions and we focus on the loop expansion of the partition function.
It is constructed using the following well-known linearization identity for $\Z_{2}$ variables:
\be
\me^{ J \sigma}
\,=\,
(\cosh J+\sigma \sinh J)\,,
\quad
\forall \sigma=\pm 1\,.
\ee
Expanding the nearest neighbor exponentials leads to a loop expansion of the Ising partition function:
\beq
Z_{\text{Ising}} 
&=&
(\cosh J)^E\,
\sum_{\{\sigma_{v}\}} \prod_{e}
(1+\sigma_{s(e)}\sigma_{t(e)}\tanh J) \nn\\
 &=&
 2^V(\cosh J)^E \,
\sum_{\gamma\in\cG} (\tanh J)^{P_\gamma}\,.
\eeq
We are summing over all even subgraphs $\gamma$ of the initial lattice $\Gamma$. These subgraphs, as said earlier, can be seen as sets of possibly intersecting loops on the square lattice. Then $P_\gamma$ is the number of edges, or equivalently the perimeter, of $\gamma$.
We proceed similarly  from \eqref{def_equa} and we define the high temperature form 
of the Ising spin network state $\ket{\psi_{\text{HT}}}$ as
\beq
\ket{\psi_{\text{HT}}} 
&=&
\left(\cosh\frac{J}{2} \right)^{\f E2}
 \sum_{\gamma\subset\Gamma}
\left( \tanh\f J2\right)^{\f {P_{\gamma}}2}
\,\ket{\gamma}
\nn\\
\ket{\gamma}
&\equiv&
\prod_{v\in\gamma}
(\hat{\tau}^z_{v})^{\eps^\gamma_{v}}
\,\ket{++\cdots}\,,
\eeq
where we have switched the intertwiner basis to:
\be
\ket{\pm} = \frac{1}{\sqrt{2}}\left( \ket{\uu} 
\pm \ket{\ud} \right)\,.
\ee
We are now summing over all possible subgraphs $\gamma$ (defines as arbitrary subsets of edges of the initial graph $\Gamma$) and not restricting ourselves to even subgraphs. The index $\eps^\gamma_{v}$ actually registered the parity of the valency of a vertex $v$ with respect to the subgraph: $\eps^\gamma_{v}=+1$ if $v$ is attached to an even number of edges of the subgraph $\gamma$, while $\eps^\gamma_{v}=-1$ if it is attached to an odd number of edges.
It is a non-trivial consistency check to show directly that the norm $\la \psi_{\text{HT}}|\psi_{\text{HT}} \ra$ is actually simply the Ising partition function.

This high temperature state is constructed not form the full spin-up state 
but from the plus state $\ket{+} = \frac{1}{\sqrt{2}}\left( \ket{\uu} 
+ \ket{\ud} \right)$ which actually sums over all possible combinations of spins up and down. This  seems natural since the high temperature regime of the Ising model is totally disordered.

From the perspective of loop quantum gravity, the $\ket{+} $ state is 
actually $\ket{0_s}$ and the $\hat{\tau}^z$ operator acting at a vertex 
$v$ is simply the normalized squared volume operator 
$\hat{\tau}^z = \frac{4}{\sqrt{3}}\hat{U}_v$. This allows to define 
this high temperature expansion of the Ising spin network entirely in 
terms of geometric operators. The interested reader will find more 
detail in appendix \ref{sum_states}.

\subsection{Loop representation}
\label{loop_representation}
\label{holodecomp}

Loop quantum gravity is based on the reformulation of general relativity as a $\SU(2)$ gauge field theory. Its basic observables, and then operators at the quantum level, are the holonomies. From the viewpoint of the spin network wave-functions, the gauge invariance is ensured by the intertwiners. It is always enlightening to understand how to reconstruct some spin network states from the basic holonomy operators and we will investigate below how to derive such a loop decomposition for our Ising spin network states.

So the Ising spin network wave-functions for a vanishing phase $\Theta[\sigma_v] =0$ reads:
\begin{align}
\label{loop_def}
\psi(g_e) = 
	\sum_{[\sigma_v]} A[\sigma_v] 
	\trace{\left( \bigotimes_e D^{1/2}(g_e) \otimes \bigotimes_v \sigma_v \right)}
\end{align}
where $ D^{1/2}(g_e)$ are the spin-$1/2$ Wigner matrix representation
of the $SU(2)$ group element $g_e$ associated to every oriented edge $e$. Changing the orientation of an edge simply simply switches $g_{e}$ to its inverse $g_{e}^{-1}$.
Using the relation between the eigenstates of the squared volume
operator and the spin zero intertwiner states in the three pairing channels,
$\ket{\uud} 
= \frac{2}{3} 
	\left( 
		\ket{0_{s}} + \me^{\mp \frac{\mi \pi}{3}} \ket{0_{t}}
		\me^{\mp \frac{\mi 2\pi}{3}} \ket{0_{u}}
	 \right)$,
%
we obtain a decomposition of the Ising states over tessellations $\mathcal{T}$ on the lattice:
\begin{align}
\label{loop_tess}
\psi(g_e) =
	\sum_{\mathcal{T}} 
	 \sum_{[\sigma_v]} A[\sigma_v] 
	\me^{\mi \frac{\pi}{3} \theta(\mathcal{T},[\sigma_v])}
	\prod_{\mathcal{L} \in \mathcal{T}} 
	\chi_{1/2} \left( \prod_{e\in\mathcal{L}} g_e \right)
\end{align}
where the phase $\theta(\mathcal{T},[\sigma_v])$ depend on both the tessellation and the Ising spins, and $\chi_{1/2}$ is the character (i.e the trace) of the fundamental representation.
We call here tessellation a set of loops covering every link of the lattice once and only once (or equivalently a partition of the edge set of the lattice in terms of closed loops).
Defining an auxiliary variable $t_v = 0, 1, 2$, as a value
for the three channels $s, t, u$  respectively, see figure \ref{mand_intert},
the phase takes a simple form,
$\theta(\mathcal{T},[\sigma_v]) \equiv \theta([t_v],[\sigma_v]) 
= - \sum_v t_v \sigma_v$.
Here, we have omitted a factor $(2/3)^{V}$ in the  normalization of the wave-function \eqref{loop_tess} coming from the decomposition of the up and down states $|\uud\ra$ in terms of the spin-0 states in the $s$, $t$ and $u$ channels.

It is interesting to note that  the amplitude associated to a given tessellation
is a partition function of a 2D Ising model with local imaginary magnetic fields. 
For a given tessellation, or equivalently channel values $t_{v}$ for every vertex, the effective Hamiltonian is
\begin{align}
H_{\text{loop}} = -\frac{J}{2} \sum_{\langle v,w \rangle} \sigma_v \sigma_w
			- \sum_v \left( B_v + \mi \frac{\pi}{3}t_v\right) \sigma_v
\end{align}
An analytical solution of the 2d Ising model with magnetic fields is actually still 
unknown. In one dimension, for $B_v = \mi h$, there exist an infinite 
number of couples $(J, h)$ for which the correlation length diverges  
\cite{Mussardo_book}. In fact, studying models with imaginary magnetic 
fields gives information on the onset of phase transitions, relaxation or 
decoherence timescales \cite{Lee_Yang_1952_I,Lee_Yang_1952_II,
Lee_Yang_decoherence}. Indeed, the Lee-Yang theorem explores the 
zeros of the partition function which in turn are related to the divergence
of thermodynamical quantities like the free energy. It would be interesting in future investigation to look at that potential interplay between the Lee-Yang zeroes and the structure of the Ising spin network states in the thermodynamical limit.

\begin{figure}[h]
  \centering
  \includegraphics{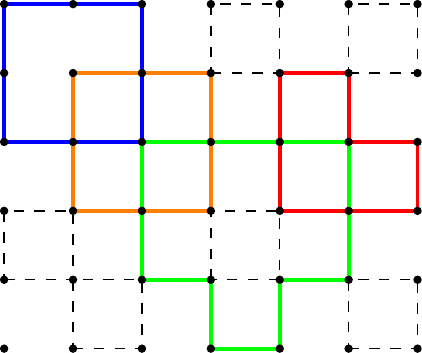}
  \caption{One example of a tessellation that appears in the loop decomposition
  		of the Ising spin network state (for clarity some loops have been 
  		omitted at the boundary). The thick loops are examples  of extended
  		loops contrary to the dashed ones. It is those that are at the origin of 
  		the non trivial correlations. }
  \label{loop}
\end{figure}

Figure \ref{loop} gives an example of a particular tessellation which 
as defined above is a set of loops passing once and only once through 
each link of the lattice. As a potential candidate for the generic 
solutions of the Hamiltonian constraint, this loop decomposition gives us 
information on the loop structure of physical states. Generically,
the emergence of non trivial correlations is a consequence of the 
quantum superposition of all possible sets of extended loops covering
the spin network.

Having the loop representation, it is interesting to check the results 
mentioned in the previous section and see that, apart from a normalization
factor, we have the same expressions of the norm and correlations. 
The method here is to write all the loop contributions as a product at 
each vertex $v$ of a matrix element $\langle \sigma'_v | M | \sigma_v \rangle$.
This matrix is quite reminiscent of the transfer matrix method used in 
statistical physics. In our case, it is proportional to the identity as
is expected from the very construction of the state. 

We give here some details on this calculation. Even if this specific 
evaluation can be done in a more straightforward way, the purpose 
here is to illustrate the method. We write the norm $\mathcal{N}$
\begin{align}
\mathcal{N}&= 
	 \sum_{\mathcal{T},\mathcal{T}'} \sum_{\{\sigma_v,\sigma'_v\}}
		\me^{\mi\frac{\pi}{3}\left( \theta(\mathcal{T},[\sigma_v]) 
		- \theta(\mathcal{T}',[\sigma'_v]) \right)}
		A^{*}[\sigma'_v] A[\sigma_v] \nonumber \\
	&\times 
	\int_{SU(2)} \prod_{\mathcal{L} \in \mathcal{T}} \chi_{1/2}
			 \left( \prod_{e\in\mathcal{L}} g_e \right)
			\prod_{\mathcal{L}' \in \mathcal{T}'} \chi_{1/2} 
			\left( \prod_{e\in\mathcal{L}'} g_e \right) 
	\nonumber \\
	&= 
	\sum_{\{\sigma_v,\sigma'_v\}}  
	\left( \sum_{\{t_v,t'_v\}}
		\me^{\mi\frac{\pi}{3}\left(\sigma' t' - \sigma t \right)}
		\langle t' | t \rangle
	\right)
	A^{*}[\sigma'_v] A[\sigma_v] 
\end{align}
with $\sigma t = \sum_v \sigma_v t_v$, 
$\langle t' | t \rangle = \langle \otimes_v t'_v | \otimes_v t_v \rangle =
\prod_v \langle t'_v | t_v \rangle $.
We then basically define the matrix $M$ as follows 

\begin{align}
\prod_v \langle \sigma'_v | M | \sigma_v \rangle = 
	\left( \sum_{\{t_v,t'_v\}}
	\me^{\mi\frac{\pi}{3}\left(\sigma' t' - \sigma t \right)}
	\langle t' | t \rangle
\right)
\end{align}
This matrix $M$ is obtained by multiplying a phase matrix $P$ and an 
intertwiner matrix $I$ whose elements are simply the different 
overlaps $\langle t'_{v} | t_{v} \rangle$, $M= {}^t\! P I P$, such that  
\begin{align}
\label{IP}
I = 
\begin{pmatrix}
1 & 1/2 & -1/2 \\
1/2 & 1 & 1/2 \\
-1/2 & 1/2 & 1
\end{pmatrix}
\quad 
P = 
\begin{pmatrix}
1 & 1 \\
\me^{-\mi\frac{2\pi}{3}} & \me^{\mi\frac{\pi}{3}} \\
\me^{-\mi\frac{2\pi}{3}}  & \me^{-\mi\frac{2\pi}{3}} 
\end{pmatrix} 
\end{align}
Doing this calculation gives $M = \frac{9}{2}\mathbbm{1}$ where we 
conclude that the norm is proportional to the Ising partition function 
(the proportionality factor being unimportant for the correlations we
are interested about),

\begin{align}
\mathcal{N} = 
	\left(\frac{9}{2}\right)^{n_v}
 	\sum_{[\sigma_v]} |A[\sigma_v]|^2
\end{align}
The same computation gives the expected correlation functions for the
intertwiner spins. 

\subsection{Density}

With this loop decomposition of the Ising  spin networks at hand, we can look at the structure of its corresponding density $\rho(g_{e})=|\psi(g_e)|^2$, which defines its probability profile. 

Squaring the Ising wave-function $\psi(g_e)$ dressed with spins $1/2$ on all edges of the lattice, we get tensor products of two spins $1/2$ on each edge, which recouple to a spin 0 or a spin 1. Therefore the probability density $\rho(g_{e})=|\psi(g_e)|^2$ will once again decompose as a gauge-invariant function on $\SU(2)^{E}$ as a superposition of spin networks dressed with spins 0 or 1 on the edges and admissible intertwiners at the vertices. 

The spin configuration  consisting in only spins 0 everywhere on all edges actually corresponds directly to the calculation of the norm $\cN$ of the density computed above. Up to a normalization factor, we have shown that we recover the Ising partition function.
We generalize this method to compute the probability of the spin configurations, which also involve some spins 1, and we will show how they are related to correlation functions of the Ising model.

\begin{figure}[h]
  \centering
  \includegraphics{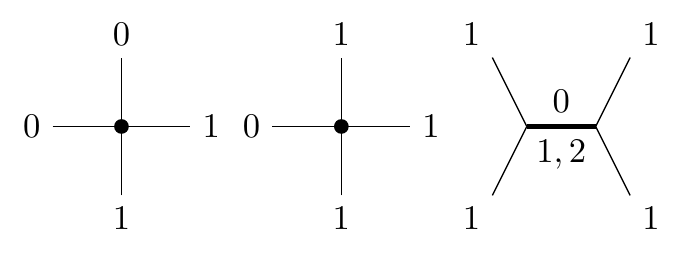}
  \caption{Graphical representation of the spin $1$ intertwiners which 
		are used in to extract parallel transport information from the 
		density.}
  \label{intert_spin1}
\end{figure}

Using the same notation as previously, namely $\ket{t}$ which 
represented the tensor product of a particular configuration of 
intertwiner states depending on some given values $t_{v}$, we can consider the probability density $\rho(g_{e})$ as a spin network state itself.
\begin{align}
\ket{\rho(g_e)} = 
	\sum_{\{\sigma_v,\sigma'_v,t_v,t'_v\}}
		A^{*}[\sigma'_v] A[\sigma_v] \,
		\me^{\mi\frac{\pi}{3}\left(\sigma' t' - \sigma t \right)}\,
		\ket{t,t'}
\end{align}
We would like to compute the overlap between the state $\ket{\rho(g_e)}$
and the state $|\cS\ra$ associated to a chosen configuration $\cS$ of spins and intertwiners:
\begin{align}
	\label{spinoneIsing}
\langle \cS | \rho(g_e) \rangle = 
	\sum_{\sigma,\sigma',t,t'}
		A^{*}[\sigma'_v] A[\sigma_v] 
		\me^{\mi\frac{\pi}{3}\left(\sigma' t' - \sigma t \right)}
		\langle \cS | t,t' \rangle
\end{align}
This defines a  transfer matrix $M^{\cS}$ between the Ising spins $\sigma_{v}$ and $\sigma'_{v}$:
\be
\la \otimes_{v }\sigma'_{v} | M^{\cS}| \otimes_{v }\sigma'_{v}\ra
=
 \sum_{\{t_v,t'_v\}}
		\me^{\mi\frac{\pi}{3}\left(\sigma' t' - \sigma t \right)}
		\langle \cS | t,t' \rangle
\ee
We cut the spin configuration $\cS$ into little pieces $\cS_{v}$ defining the spin configurations around each vertex $v$. The various cases are given on fig.\ref{intert_spin1}. Then we realize that the transfer matrix $M^{\cS}$  can be factorized vertex by vertex:
\begin{align}
\la \otimes_{v }\sigma'_{v} | M^{\cS}| \otimes_{v }\sigma'_{v}\ra
=
\prod_v \langle \sigma'_v | M ^{\cS}_{v} | \sigma_v \rangle \\
\textrm{with}\quad
\langle \sigma'_v | M ^{\cS}_{v} | \sigma_v \rangle
=
\me^{\mi\frac{\pi}{3}\left(\sigma'_{v} t'_{v} - \sigma_{v} t_{v} \right)}
		\langle \cS_{v} | t_{v},t'_{v} \rangle\,.
\end{align}
Omitting the unambiguous vertex index $v$, we proceed as in the previous section and write the transfer matrix as $M^{\cS} = ^t\! P I^{\cS} P$ where $P$ is the phase matrix defined in the previous section in eqn.\eqref{IP}
and $I^{\cS}$ the $3\times 3$ matrix interpolating between  $t_{v}$ and $t'_{v}$  and whose elements are the scalar products $\langle \cS_{v} | t_{v},t'_{v} \rangle$.

The task is then to evaluate those matrices.
On the one hand, we have the states $\ket{t,t'}$, which are simply 
tensor products of the states $|0,1\ra_{s}$ and in the two other channels as defined in section \ref{def}, eqn.\eqref{intert_state_12}.
On the other hand, we have the spins 0 and 1 and we need the intertwiners between spins 1. There are three different cases, whether we recouple between 2,3 or 4 spin-1 at the considered vertex, as illustrated on figure \ref{intert_spin1}. The bivalent and trivalent cases define a unique intertwiner. The four-valent is more involved since the intertwiner space is 3-dimensional.

For the purpose of clarity, we give explicit details on a particular 
example. The other cases can be treated similarly. We focus on the case with spin $1$ on the third and fourth edges, as the first case of figure \ref{intert_spin1}. This intertwiner, denoted  
$\ket{I_{0011}}$, is equal to $\ket{0_{12}} \ket{1_{34}}$ , with 

\begin{align*}
\ket{0_{12}} &= 
	\frac{1}{\sqrt{2}}
	\left( 
	\ket{\begin{array}{ c }
			\uparrow\\
			\downarrow
		\end{array}} -
		\ket{\begin{array}{ c }
			\downarrow  \\
			 \uparrow
		\end{array}}
	\right)
\otimes
\frac{1}{\sqrt{2}}
	\left( 
	\ket{\begin{array}{ c }
			\uparrow\\
			\downarrow
		\end{array}} -
		\ket{\begin{array}{ c }
			\downarrow  \\
			 \uparrow
		\end{array}}
	\right)\\
\ket{1_{34}} &= 
\frac{1}{\sqrt{3}}
	\left[
		\ket{\begin{array}{ c c }
			\uparrow & \downarrow \\
			\uparrow & \downarrow
		\end{array}}	 +
		\ket{\begin{array}{ c c }
			\downarrow & \uparrow  \\
			\downarrow & \uparrow
		\end{array}} -
		\frac{1}{2}
			\left(
			\ket{\begin{array}{ c }
				\uparrow \\
				\downarrow 
			\end{array}} +
			\ket{\begin{array}{ c }	
				\downarrow  \\
				\uparrow 
			\end{array}}
			\right)
	\right]
\end{align*}
The column notation mirrors the presence of two copies of spin $1/2$, 
the top line representing the state one copy and the bottom one the other.
The matrix $I_{0011}$ then consists in the scalar product $\la 0_{12}1_{34}| 0_{s,t,u}0_{s,t,u}\ra$.
Calculating those overlaps is now straightforward. The interested reader will find the explicit expressions of the states $|0_{s,t,u}\ra$  in appendix. We obtain 
the intertwiner matrix $I_{0011}$:
\begin{align}
I_{0011} =  -\frac{\sqrt{3}}{4}
 	\begin{pmatrix}
		1 & 1/2 & -1/2 \\
		1/2 & 0 & -1/2 \\
		-1/2 & -1/2 & 0
	\end{pmatrix}
\end{align}
The last step is to perform the multiplication by the phase matrix $P$ 
to finally have the specific $2\times 2$ transfer matrix $M_{0011}$, 
\begin{align}
M_{0011} &= -\frac{3\sqrt{3}}{4} 
		    \left(  \frac{1}{2}\ \mathbbm{1}  + \hat{\tau}_x \right)\,.
\end{align}
Let us remember that this matrix acts as an operator insertion between the Ising spins $\sigma_{v}$ and $\sigma'_{v}$. Looking at the probability amplitude $\langle \cS | \rho(g_e) \rangle$ defined in eqn.\eqref{spinoneIsing}, we will have to insert all of the transfer matrices for each vertex.
We give in appendix \ref{Matrices_intert} the complete set of transfer matrices and operators
associated to each type of configuration of spins and  intertwiners.
Having those, we can see that for a
particular configuration $\cS$ the amplitude is given by a
correlation function of the Ising model:
\be
\langle \cS | \rho(g_e) \rangle
\,=\,
\sum_{\{\sigma_{v},\sigma'_{v}\}}
A^{*}[\sigma'_v] A[\sigma_v]
\,\prod_{v}\la\sigma'_{v}|M^{\cS}_{v}|\sigma_{v}\ra\,,
\ee
where each of the operator insertion $\la\sigma'_{v}|M^{\cS}_{v}|\sigma_{v}\ra$ will consist in a Pauli matrix combination, switching or phasing the Ising spins at every vertex $v$.


This shows how to express the probability density defined by the Ising spin network in terms of the Ising correlation functions.

\section{Hamiltonian constraints}
\label{hamil}

\subsection{Definition and algebra}
\label{hamiltonian}

Up to now, we have focused on defining Ising spin network states such that their $n$-point functions map onto the correlations of the standard Ising model and on how to generate them using geometric and holonomy operators, but they remain kinematical states of loop quantum gravity. It would be more interesting and physically relevant if we could build physical states solving the Hamiltonian constraint operators or if we could interpret our Ising spin networks as at least approximate physical states in some regime. Indeed, the Hamiltonian constraints are crucial in loop quantum gravity, they generate the dynamics of the theory by implementing its invariance under space-time diffeomorphism and are the quantum Einstein equations for quantum gravity. However there isn't yet a systematic method to determine physical states, or at least a perturbative approximation scheme, in loop quantum gravity despite several recent lines of research (especially using spinfoam models \cite{Perez:2012wv}) since Thiemann's original proposal \cite{Thiemann_1996} (see e.g. \cite{Bonzom:2011jv} for a review of the various approaches through toy models).

Here we will build some local Hamiltonian constraint operators using the basic geometric operators of loop quantum gravity, such that the Ising spin network states are their only unique solutions. Of course, these do not have any a priori link with actual gravity or any proposal of dynamics for loop quantum gravity. Moreover, we are working on a fixed graph and one might expect the quantum gravity dynamics to act on the spin network graph. Nevertheless, despite all these shortcomings, there might be some lessons to draw for such a toy construction. 

First it illustrates the type of Hamiltonian constraints that would lead to Ising-like spin network states and thus to admitting non-trivial long range correlations in a continuum limit. Second we will build our constraint operators from the volume operators acting on nearest neighbor vertices, or equivalently around loop on the dual lattice. This suggest a change of perspective from the usual methods to construct regularized Hamiltonian constraint operators in loop quantum gravity, which focus on the holonomy operators around loops on the actual lattice.

\smallskip

Let us thus construct some Hamiltonian constraint operators characterizing the Ising spin networks.
From the condensed matter perspective, a first approach could be to
look at the parent Hamiltonian technique and build a Hamiltonian
operator whose ground state is our considered Ising state 
\cite{Cirac_2006}.
We will follow a simpler path, looking at the action of the basic Pauli matrix operators on the Ising spins and using them to define simple local Hermitian Hamiltonian constraint operators characterizing uniquely our Ising 
states.

Let us consider the Pauli operator $\htau_{v}^{x,y,z}$ acting on the Ising spin at the vertex $v$ and first focus on the switching operator  $\htau^{x}\,|\uu\ra=|\ud\ra$, $\htau^{x}\,|\ud\ra=|\uu\ra$.
Its action on the Ising state can be re-absorbed in the amplitude:
\begin{align}
\htau_v^x\,\ket{\psi} &=
	 	\sum_{\{\sigma_i\}}  
		A[\sigma_{i}]\,
		\ket{-\sigma_v}\otimes\bigotimes_{i\ne v} \ket{\sigma_i} \\
&=  \sum_{\{\sigma_i\}} 
		A[\sigma_{i}]\,
		\me^{-J\sum_{\langle v,w \rangle} \sigma_v \sigma_{w}}\,
		\me^{-B_{v}\sigma_{v}}\,
		\bigotimes_{i} \ket{\sigma_i} \nn\\
\textrm{with}&
\quad
A[\sigma_{i}]
= \me^{\frac{J}{2}\sum_{\langle i,j \rangle} \sigma_i \sigma_{j} +\f12\sum_{i}B_{i}\sigma_{i}}\,. \nn
\end{align}
Up to the magnetic field term, the change of sign of the Ising spin at the vertex $v$ translates into the  insertion of an 
additional factor involving its nearest neighbor spins,
$\me^{-J \sum_{\langle v,w \rangle} \sigma_w \sigma_v }$.
We turn this factor in its linearized form, 
$\me^{-J \sigma_v \sigma_w}= \cosh(J) - \sigma_v \sigma_w \sinh( J)$,
which allows us to define the following Hamiltonian constraint operator at the vertex $v$:
\begin{align}
\hat{H}^x_v = \htau_v^x - 
			\hat{h}_{v} 
			\prod_{\langle v, w \rangle}  \hat{h}_{vw}
\end{align}
where we use for simplicity the auxiliary operators
\footnote{
These auxiliary operators are easily invertible:
\beq
\hat{h}_{v}^{-1} &=& \cosh B_v + \htau^z_v \sinh B_v\nn\\
\hat{h}_{vw}^{-1} &=& \cosh J + \htau^z_v \htau^z_w \sinh J\nn
\eeq
}
:
\beq
\hat{h}_{v} &=& \cosh B_v - \htau^z_v \sinh B_v\\
\hat{h}_{vw} &=& \cosh J - \htau^z_v \htau^z_w \sinh J
\eeq
We have just showed that these constraints annihilate our Ising spin network state:
\be
\hat{H}^x_v 
\,\ket{\psi}
\,=\, 0\,.
\ee
Let us remember that the operator $\htau^z_v$ is simply the normalized square volume operator acting at the vertex $v$.

Similarly, the action of the other Pauli matrix operator $\htau_{v}^{y}$ leads us to  define another set of Hamiltonian constraint operators $\hat{H}^y_v$:
\begin{align}
\hat{H}^y_v = \htau^y_v + \mi \htau^z_v 
			\hat{h}_{v} 
			\prod_{\langle v, w \rangle} \hat{h}_{vw}
		= -\mi \htau^z_v \hat{H}^x_v 
\end{align} 

\begin{figure}[h]
  \centering
  \includegraphics{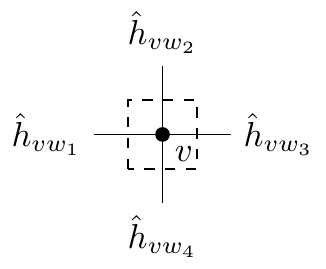}
  \caption{A nearest neighbor vertex operator can be seen as a loop
		operator on the dual lattice around the vertex $v$.}
  \label{vertex_loop}
\end{figure}

Since the operators $\htau^{x,y,z}_{v}$ can all be constructed from the area and volume operators, those Hamiltonian constraints are entirely expressed in terms of geometric operators.
This differs slightly from the usual construction of the Hamiltonian constraint operators in loop quantum gravity, based on Thiemann's trick \cite{Thiemann_1996} and which involves the holonomy operator (as regularization for the curvature) around loops of the spin network graph.
Here, the product of operators $\prod_{\langle v, w \rangle} \hat{h}_{vw}$ over the nearest neighbors of the vertex $v$ can be also considered as living on a loop but on the dual lattice, as illustrated on fig.\ref{vertex_loop}. We will discuss more this shift of perspective below in section \ref{insight}.

We note here that the possible local 
spin dependent phases that could be added in the
definition of the state (Eq.\ref{def_equa}) which do not change the correlation
functions would change the Hamiltonian constraints previously defined.

By construction, the Ising spin network state is solution of all these constraints, 
$\hat{H}_v^{x,y}\ket{\psi} = 0$.
To check that there are not any other constraints, we check  the algebra
they generate. The constraints for two vertices which are not nearest neighbor vanish.
The commutation relations on a single site $v$ or
for nearest neighbors  $\langle v, w \rangle$ read:
\begin{align}
\label{algebra}
\left[ \hat{H}^x_v,\hat{H}^y_v  \right] &=  
	\htau^{[x}_v\hat{H}^{y]}_v
	\\
\left[ \hat{H}^x_v,\hat{H}^x_w  \right] &=  
	2 \sinh J\,\, \hat{h}_{vw}  \htau^y_{[v}\hat{H}^y_{w]}
\nonumber \\
\left[ \hat{H}^y_v,\hat{H}^y_w  \right] &=  
	2 \sinh J\,\, \hat{h}_{vw}  \htau^x_{[v}\hat{H}^x_{w]}
\nonumber \\
\left[ \hat{H}^x_v,\hat{H}^y_w  \right] &=  
	2 \sinh J\,\, \hat{h}_{vw}   \htau_{[v}^{y}\hat{H}_{w]}^{x}
	\nonumber 
\end{align}
Every other commutation relations are zero. So the constraint algebra does not generate any further constraints satisfied by our Ising spin network states.

\smallskip

Our Hamiltonian constraints $\hat{H}_v$ exist for  all values of the Ising coupling $J$. They do not depend on the specific phase of the Ising model, it doesn't see a priori the structure of the correlations or the phase transition.
We think that this setting would be the perfect testing ground for any coarse-graining scheme for spin network states in loop quantum gravity (e.g. \cite{Livine_2014,Livine_Terno_2005,Dittrich:2013bza,Dittrich:2014ala,Dittrich:2014wda}). Indeed the phase diagram, coarse-graining and phase transition of the 2d Ising model is entirely under control and we know what to expect from the coarse-graining flow. The coarse-graning procedure for loop quantum gravity should reproduce the same flow on the correlations of the Ising spin network states. It will then be enlightening to understand what happens at the level of the Hamiltonian constraint operators, how the emergence of large scale correlations is taken into account and what triggers or signals the phase transition at the critical Ising coupling.

Another approach would be to have  Hamiltonian constraints specifically tuned to the coarse-graining properties of the Ising partition
function, which would not determine the Ising spin network states for arbitrary values of the coupling $J$ but that would select
 specifically the critical point. Such Hamiltonian operator would reflect the exact coarse-graining flow of the Ising model. This would be much more complicated to realize than our present proposal. It would either involve a graph changing dynamics or implement a self-duality property of the Ising partition function at the
critical temperature, probably through the square-star or triangle-star 
relations \cite{Baxter_book}. We believe that this is a very interesting line of research, but yet out of the scope of the present work and we postpone it to future investigation.
 
\subsection{Unicity}

So far, we have introduced a set of constraints $\hat{H}_v^{x,y}\ket{\psi} = 0$ of which the Ising spin network
state is a solution. We  now show that is it the unique
solution of those constraints (up to a global phase factor). 

The state of the intertwiners has the following general form
\begin{align}
\ket{\psi} = \sum_{[\sigma_v]}
		\alpha_{[\sigma_v]} \bigotimes_v \ket{\sigma_v}
\end{align}
where $\alpha_{[\sigma_v]}$ are the coefficients of the state on the intertwiner basis. Imposing that the state solves the constraints $\hat{H}_v^{x,y}\ket{\psi} = 0$ leads us to a detailed balance type 
condition between a configuration and another with only one spin 
flip difference at a vertex $v$,
\begin{align}
\alpha_{\sigma_1, \cdots, -\sigma_v, \cdots, \sigma_N} = 
\me^{-J\sum_{\langle v,w \rangle} \sigma_v \sigma_w}
\alpha_{\sigma_1, \cdots, \sigma_v, \cdots, \sigma_N}
\end{align}
The solution of this relation is found as follows. Beginning
at a particular configuration, namely the one with all spins
up $\uu$, we get the amplitude of an arbitrary configuration
by flipping the relevant spins in that balance equation. The exponential factor that appears
only depends on the number of pairs of anti-parallele spins.
Using the low temperature expansion notation of 
section \ref{ht_lt} with down spin clusters $C$, we obtain that 
\begin{align}
\alpha_{[\sigma_v]} &= \alpha \me^{-J P_{C} }\,,
\qquad
|\alpha|^2\left(\sum_{C}\me^{-2J P_{C} } \right) = 1 
\end{align}
where  the amplitude  $\alpha$ associated to the totally ordered configuration  with only $\uu$ spins
 is fixed by the normalization condition . We recognize the low 
temperature expansion  of the Ising partition function. So we conclude
that the amplitude is then proportional to the Boltzmann factor
of the nearest neighbors Ising model with coupling $J/2$. Thus the 
constraints have a unique solution up to a global phase given by the 
Ising spin network state. We note that because of the relation between
the  constraints $\hat{H}_v^{x}$ and $\hat{H}_v^{y}$, we only require either one to reach 
this conclusion. For more complex qubit models, this might not be the case.

This discussion also enlightens the action of the Hamiltonian constraints and 
the dynamics they could create. Their role is to impose a detailed balance 
condition between two different configurations of intertwiners. Such behavior
is often encountered when studying stochastic systems relaxing toward equilibrium
such as for instance Glauber dynamics for Ising models which consists exactly
at looking at local spin flips. 

\subsection{Insight for Loop Quantum Gravity}
\label{insight}

The primary purpose of our Ising spin network states is to provide a toy model framework to study coarse-graining in loop quantum gravity. Working in the controlled environment of the Ising model with its explicitly known correlations seems to be a perfect testing ground to investigate coarse-graining procedures, continuum limit definitions and phases transitions.
But beyond this aspect, it turns out that they could also bring some insight into the structure of the dynamics of loop quantum gravity.

Indeed, as we have pointed out earlier in section \ref{hamiltonian}, the Hamiltonian constraint operators that we introduce for the Ising states are different from the typical construction in loop quantum gravity, based on holonomy operators around loops of the spin network graph or creating such loops \cite{Thiemann_1996,Brunnemann:2004xi} (also see \cite{Livine_Tambornino_2013} for a more recent reformulation of the holonomy and grasping operators in spinorial terms). These holonomies usually enter the constraints as a regularization of components of the curvature tensor. Here the natural structure of our Hamiltonian operators involve volume operators on dual loops (living on the dual graph) acting on all nearest neighbors of a given vertex. Although our present construction has clearly no a priori link with gravity, it seems closer to the intuition of gravitational waves deforming volumes and shapes from vertices to vertices. This suggests to look for a reformulation of the  loop quantum gravity dynamics in such terms, or more generally to investigate the relation (e.g. under the form of a dual expansion) between the two types of Hamiltonian. This might help seeing gravitation waves (or at least geometry deformation waves) emerge in loop quantum gravity directly at the level of the Hamiltonian constraint algebra and not only in  a large scale semi-classical limit.

From our viewpoint, this would require, first un-freezing the spins on the spin network edges to allow from holonomy operators to shift those spins, second to work out how our Hamiltonian constraints for the Ising spin network states can be re-written as some fixed-spin projection of more general operators built of holonomies. To this purpose, our decomposition of the Ising states in terms of basic holonomies given in section \ref{holodecomp} might be a good starting point. Then we hope to generalize this discussion to the full loop quantum gravity framework outside the very restricted toy model of the present Ising spin network states.

\section{Phase diagrams and continuum limit}
\label{phase_diag}

\subsection{Orientation alignment}

\begin{figure}[h]
  \centering
  \includegraphics{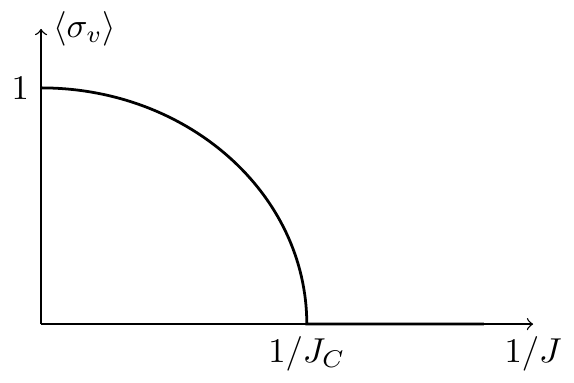}
  \caption{Generic representation of the behavior of the magnetization 
  		(in normalized units) as a function of the coupling passing
  		through a second order phase transition.}
  \label{aimantequation}
\end{figure}

The effective spins we used were defined using the two eigenstates
of the square volume operator which physically correspond to two 
different possible orientations of the fundamental volume. It is interesting to identify a regime in which all orientations would be aligned, for instance either all in the positive sector of the $\hat{U}_{v}$ operators or in their negative sector. This is exactly what happens for the Ising model, in both 2d and 3d, at low temperature or equivalently at high enough coupling $J$ in our setting. Indeed beyond the critical coupling, 
$J_c  = \frac{\ln(1 + \sqrt{2})}{2}$  for the regular square lattice  \cite{Baxter_book}, the Ising mode predicts an ordered state, with all the Ising spins align with each other. More precisely,  a phase transition occurs for the Ising model from a disordered to an ordered state.
Figure \ref{aimantequation} shows the generic behavior of the
magnetization $\langle \sigma_v \rangle$ as a function of the coupling 
$J$ (see eqn.\eqref{exactvalues} for exact formulas in the 2d case).
Above the critical coupling, the system acquires an average
orientation direction, randomly picked when passing the phase transition.
Of course  the system is perfectly ordered only in the infinite coupling limit (zero temperature 
in statistical physics) and gets quantum fluctuations away from it.
%

As for the Ising spin network states, in the infinite coupling limit, the state reduces to the sum over the two limit states, the one where the intertwiners are all in the positive volume orientation state plus the one where the intertwiners are all in the negative volume orientation state. As the coupling decreases,  opposite spin clusters appear and their typical size increases with the correlation lengths. Towards the phase transition, we have clusters of all sizes with no apparent ordering or preferred orientation. Beyond the phase transition, at lower coupling, we are in the disordered phase, with no orientation alignment. An ordered universe, with aligned volume orientation, would therefore live in the ordered phase at high coupling $J$ (i.e low temperature).
%
%

\subsection{Distance from correlations}

In a non-perturbative background-independent approach to quantum gravity, the very notions of 
distance, locality or metric need to be reconstructed from scratch in the absence of a
background geometry. Understanding the emergent geometry from 
the quantum state is still an ongoing issue and we hope to be able to recover all this geometric information from the correlations and entanglement in the quantum state \cite{Livine_Terno_2005}.

So the idea we pursue here is that a notion of distance 
emerges from the correlations in the spin network state. Using the known results on the Ising model, we get the Ising spin network correlations in terms of the natural lattice distance and then work backwards attempting to see to what extent the distance between two spins can get be extracted from the sole knowledge of the correlations between those spins.
By construction, the Ising spin networks have exactly the same known correlation
behavior than the classical Ising model, namely
\begin{align}
\langle \sigma_ i \sigma_j \rangle  -\langle \sigma_i \rangle\langle \sigma_j \rangle   \propto 
	\left\{ 
		\begin{array}{ c c }
			 C\exp{\frac{-\,|i-j|}{\xi}}, & J \ne J_c \\
			\frac{1}{|i-j|^{1/4}}, & J=J_c
		\end{array}
	\right.
\end{align}
where
$\xi$ is the correlation length (in units 
of the lattice spacing) and $C$ a positive constant. 
%
Here the distance between the vertices $|i-j|$ is the graph distance as naturally defined on the 2d regular square lattice.
The exact form of 
the magnetization and the behavior of the correlation length near 
the transition are known but won't be used in the following
\footnote{For reference, below the critical coupling
\begin{align}
\label{exactvalues}
\langle \sigma_v \rangle^2 &= \left[ 1 - 
						\left(\frac{1-(\tanh{J})^2}{2(\tanh{J})^2}\right) 
				     \right]^{1/4}  \nonumber \\
\xi &\propto \left|\frac{1}{J} - \frac{1}{J_c}\right|^{-1}
\end{align}}.

\begin{figure}[h]
  \centering
  \includegraphics{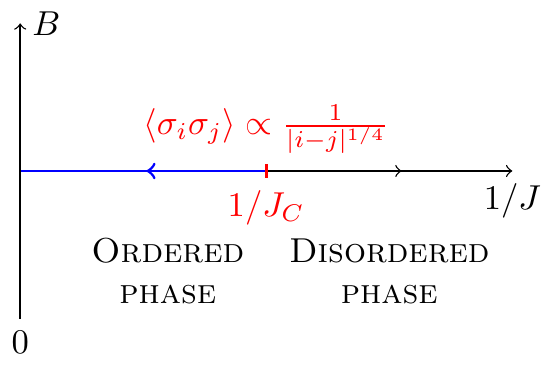}
  \caption{Phase diagram with renormalization flow of the Ising spin 
		network state with the algebraic decaying correlation 
		function at the critical coupling.}
  \label{phasediag}
\end{figure}

It is possible to invert the correlation formula so as to express the distance as a function of the correlation itself. The distance will be generically a monotonous decreasing function of the proper correlation between the two spins, $\langle \sigma_ i \sigma_j \rangle_{p}\equiv\langle \sigma_ i \sigma_j \rangle  -\langle \sigma_i \rangle\langle \sigma_j \rangle $, but the precise function will depend on the considered phase and regime.
Away from the critical point, we have an exponential decay of the correlation and we can define the distance as $d(i,j)\equiv -\xi\,\log \langle \sigma_ i \sigma_j \rangle_{p}$ up to an additive constant. The correlation scale $\xi$ becomes the new length unit. At the critical point, things become especially interesting. The correlation length blows to infinity and we have an algebraic  decay of correlation. We can now define the distance as $d(i,j)\equiv 1/ \langle \sigma_ i \sigma_j \rangle_{p}^{4}$.

From the point of view of gravity, the most interesting case is clearly  an algebraic 
decay of the correlations. Ideally, we'd like to derive a quadratic decay of correlation, just as in standard quantum field theory, in order to retrieve Newton's gravity law. In order to get this, we expect of course to have to change our Ising model state to another better suited statistical model, but also move to a 3d lattice or graph structures, unfreeze the spins of the spin network and so on.  Nevertheless, the present Ising spin network state allows to illustrate a couple of important points:
\begin{itemize}
\item Reconstructing the distance from the (2-point) correlation depends not only of the considered statistical model on the specific phase of that model.

\item We naturally have a algebraic decay of the correlation in terms of the distance, and vice-versa, at the phase transition. At that point, the state also admits a non-trivial continuum limit.

\end{itemize}

Finally, in order to truly validate this notion of reconstructed distance, we need in general to check the triangle inequality satisfied by the distances between three points, which would involve the 3-point fluctuations. Here we know that the reconstructed distance is the initial graph distance, which satisfies this requirement. But if we would consider another spin network state, not necessarily related to a known local statistical model, this would have to be checked.

%
%

\subsection{Coarse-graining}

One of the main challenge of the loop quantum gravity program
is to define and understand its renormalization flow, from the Planck scale to large scales and derive the low energy behavior 
as semi-classical general relativity with a systematic method to compute the perturbative quantum corrections.
To this purpose, coarse graining procedures for the spin network states and the dynamics of the theory are essential tools \cite{Livine_Terno_2005, Dittrich:2013bza, Livine_2014, Dittrich:2014ala}.

Understanding coarse-graining in loop quantum gravity is not
an easy task because of  the absence of a background geometry or structure and the complicated nature of the Hamiltonian constraints.
%
%
In the present framework of the Ising spin network state, we have fixed a background graph, to a regular 2d square lattice up to now (we will deal with its three-dimensional generalization below in the next section \ref{3d_case}), and we can use this structure to define the coarse-graining flow as in standard statistical physics and condensed matter models. Thus this provides a neat toy model to test all coarse-graining procedures of loop quantum gravity.

\smallskip

In statistical and condensed matter physics, renormalization 
group and coarse graining methods \cite{Goldenfeld_book} are
widely used to understand critical phenomena. We wish to import these methods and results to quantum gravity to understand better the emergence of critical regimes.
The usual formulation of loop quantum gravity focuses on the Hamiltonian constraints, their algebra and their flow generating the time evolution. Here, for our Ising spin network states, the Hamiltonian constraints that we introduced do not see a priori the phase transition:  nothing obvious 
happens at the level of the algebra of the constraints when the 
coupling reaches its critical value.
The information about the critical regime and phase transition is truly in the coarse-graining flow. We could, for example, apply the tensor network renormalization tools to our Hamiltonian constraints, such as it was done for the so-called spin-net toy models in \cite{Dittrich:2013bza}, in order to derive their coarse-graining flow and check if it fits as expected the known flow of the 2d Ising model.

\smallskip

\begin{figure}[h]
  \centering
  \includegraphics{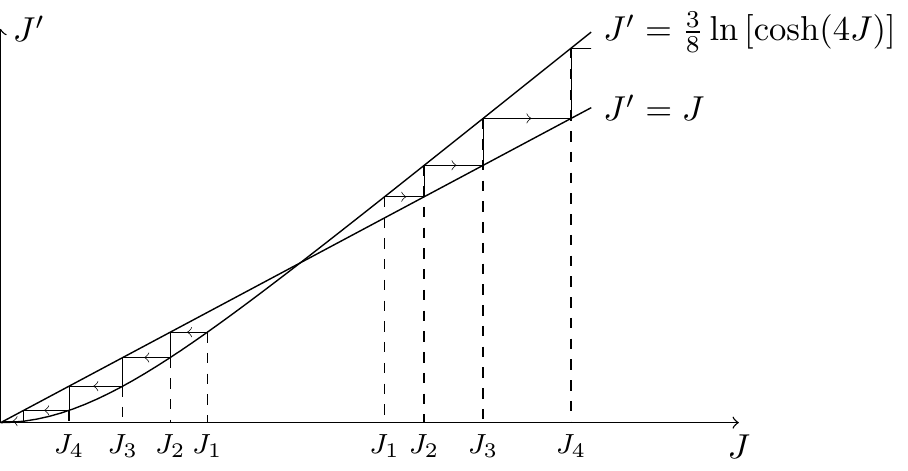}
  \caption{Renormalization group flow for the 2d Ising model using 
  		the method of decimation. The critical point is totally 
  		repulsive.}
  \label{renormequation2d}
\end{figure}
Since the spin correlations of the Ising spin network states are the same as the Ising model, we know their coarse-graining flow from the standard techniques in statistical physics. For instance, Figure \ref{renormequation2d} show its renormalization group flow using the basic isotropic decimation method:  we plot  the effective coupling $J'$ at larger scale (including the nearest and  next nearest neighbor interactions) as a function of the smaller scale coupling  $J$.  We can follow the renormalization flow of the effective coupling by iterating this map and the critical point is identified as its non-trivial fixed point.
In the critical regime, the state presents scale invariance which renders 
irrelevant the micro structure of the theory and allows for a well-defined continuum limit.
However, this fixed point is repulsive for the Ising model: as we coarse-grain, the effective coupling runs aways from the critical regime towards either the low or high temperature fixed points. We hope to later identify a better statistical physics model such that the coarse-graining flow would run instead under the coarse-graining flow towards the critical regime with algebraic decaying correlation and continuum limit. The corresponding spin network state, built along the same lines as our Ising spin networks, would then be better suited to the definition of the continuum limit and semi-classical regime of loop quantum gravity.

\subsection{Going tridimensional}
\label{3d_case}

\begin{figure}[h]
  \centering
  \includegraphics{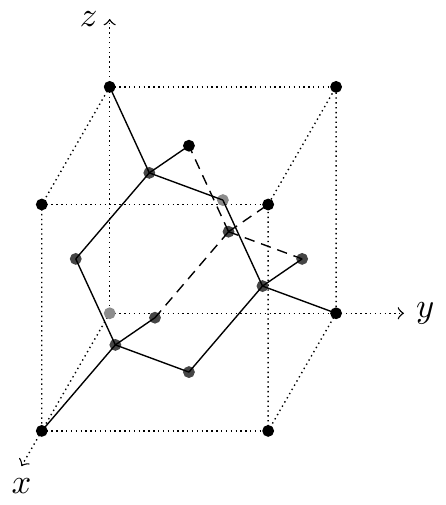}
  \caption{Elementary cell of the diamond lattice which is the natural 
		one in 3d in loop quantum gravity for a 4-valent spin 
		network.}
  \label{diamond}
\end{figure}

So far, we restrain the discussion of the gravitational state to a 
two dimensional lattice for exposition simplicity but it happens that
the generalization to 3d is straightforward.
Keeping the requirement that the lattice be 4-valent (so as to keep a two-dimensional intertwiner space and the map to effective qubits), the natural regular lattice is the diamond lattice  as illustrated on Fig.\ref{diamond}. Under the usual geometrical interpretation of loop quantum gravity, this lattice can be seen as dual to a triangulation of the 3d space in terms of tetrahedra dual to each vertex. This can be seen as an extension of the more used cubic lattice better suited to loop quantum gravity.
The Ising spin network state and the whole 
set of results which followed are then identical : the wave function 
eqn.\eqref{loop_tess}, the Hamiltonian constraints and their algebra 
eqn.\eqref{algebra} are the same.

\begin{figure}[h]
  \centering
  \includegraphics{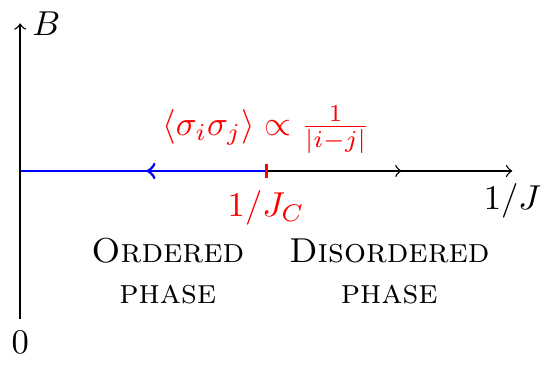}
  \caption{The 3d phase diagram of the spin network state on the 
		honeycomb lattice has a phase transition whose correlations
		behave with a one over the separation law.}
  \label{phasediag3D}
\end{figure}

In 3d, the Ising model also exhibits a phase transition. Even if the exact
solution is still unknown and is the subject of active research, much can be 
learn of its behavior around the critical coupling. Like its 2d counterpart,
the Ising spin network state has an ordered phase which corresponds
to an orientation alignment of the elementary chunk of space. Concerning
the 2-points correlation functions, information about its long distance
behavior at the phase transition or near it can be obtained using methods 
of quantum field theory. In $d$ dimensions we have \cite{Baxter_book}

\begin{align}
\langle \sigma_ i \sigma_j \rangle  \propto
	\frac{1}{(2\pi)^{d-2}}
	\frac{1}{(|i-j|)^{d-2}}
	K_{\frac{d-2}{2}}\left( \frac{|i-j|}{\xi}\right)
\end{align}
where $K(r)$ are modified Bessel functions and $\xi$ is the correlation
length. We mention that the correlation does not depend  in the 
long distance regime on the "magnetic field" $B$ (even local ones $B_i$
if we choose to define the state in this way). For the three dimensional 
case, we have the simple and exact expression 

\begin{align}
\langle \sigma_ i \sigma_j \rangle  \propto
	\frac{1}{4\pi |i-j|}
	\me^{-\frac{|i-j|}{\xi}}
\end{align}
Figure \ref{phasediag3D} represents the phase diagram of the 3d Ising
spin network (in the presence of a "magnetic field"). At the critical 
coupling, the correlations have a one over the spacing power law decay
 at long distance, still not the inverse square law but less
exotic then the $2d$ behavior. Understanding distance from
correlations in 3d is much more vital than in the 2d setting since 
in 3d their exist loops that can wind around a vertex without visiting
other distant vertex (in a relational sense).

The expression for the correlation is  obtained in a mean field setting where 
only a correlation length appears. However statistical systems possess
another length scale, the lattice spacing, which leads to correction 
of this expression through the anomalous dimension. In quantum 
gravity the Planck length  plays this role and it would be interesting
to understand its influence on the behavior of the correlations.

\section{Conclusion}
\label{Conc} 

In this paper, we have introduced a class of spin network states for loop quantum gravity on 4-valent graph. Such 4-valent graph allows for a natural geometrical interpretation in terms of quantum tetrahedra glued together into a 3d triangulation of space, but it also allows us to map the degrees of freedom of those states to effective qubits. Then we were able to define spin network states corresponding to known statistical spin models, such as the Ising model, so that the correlations living on the spin network are exactly the same as those models. For instance, we defined some Ising spin network states, both in 2d and 3d, on 4-valent regular lattice. In 2d, we use the usual square lattice, while the 3d case used the diamond lattice.

We want to propose those Ising spin network states and their generalizations (for instance to the XY model, Potts models and others) as toy models to test coarse-graining methods in loop quantum gravity and allow for explicit discussions on phase diagrams and transitions. As a start, we used them to illustrate the relation between correlations (2-point functions to be precise) and distances. Indeed, in a background independent framework with no a priori geometrical structure, all  the geometry needs to be reconstructed from the quantum gravitational state and can only emerge from its correlation and entanglement structure. Here, using the exact phase diagram and correlation functions of the Ising model, we showed how the distance could be defined a posteriori as a monotonous decreasing function of the correlation, following the intuition that two systems are close if they interact much and consequently are strongly correlated. Now the explicit map between correlation and distance actually depend on the considered phase and regime: in the ordered or disordering phase, the distance would be minus the logarithm of the correlation, while we  get an algebraic decay at the critical point. More generally, we hope that it will be possible to export the statistical physics methods to study the detail of the coarse-graining and renormalization flow of those Ising spin network states.

We further described the Ising spin network states in details from the point of loop quantum gravity, defining them in terms of geometrical and holonomy operators, showing for instance that getting long-range correlations required holonomy operators around arbitrarily large loops on the graph (and not only local holonomies). We also introduced Hamiltonian constraints that characterize uniquely our Ising states. This opens the possibility of, on the one hand, investigate a potential reformulation of the actual loop gravity constraints in similar terms, and on the other hand, analyze the coarse-graining flow of such Hamiltonian constraints and algebra in the thoroughly-studied environment of the Ising model where everything is under control.

In fact, we view our construction as a first step towards studying the coarse-graining of spin 
network states and the loop quantum gravity dynamics from the condensed matter perspective. A similar angle of attack has been developed also in \cite{Bianchi_2015} towards the construction of spin network with controlled entanglement and in \cite{Dittrich:2013bza} with the application to quantum gravity toy models of statistical coarse-graining methods such as tensor network renormalization.
We see a few lines of development of the present work. First, we should deepen the analysis of the proposal of the distance reconstruction from the correlations and go beyond the 2-point correlation to the 3-point and many-point correlation to check triangular inequalities and more generally to what extent we get a metric structure. Second, we should study the coarse-graining and renormalization flow of the Hamiltonian constraints and compare it to the standard flow of the Ising model. Third, we should generalize our construction to more generic states of loop quantum gravity, allowing for more freedom in the spins and intertwiners, and therefore extending our definition of spin network states to more complex statistical physics models with more interesting critical regimes.
To conclude, we believe that our construction of spin network states as qubit statistical models offers a perfect arena as a toy model to investigate the structure of correlation and entanglement in spin networks and the emergence of phase transitions in loop quantum gravity, and we hope to use this tool to investigate further and test renormalization methods of the Hamiltonian constraint operators in that context.

\appendix

\section{On 4-valent intertwiners between spin-$1/2$}
\label{sum_states}

In the section we present a brief summary of the different states
used in the core of the discussion and their relations.  We consider the Hilbert space $\cV$ of the spin $1/2$ representation of $\SU(2)$. The natural basis of this two-dimensional space  is denoted $\ket{\uparrow,\downarrow}$.
We now look at the intertwiner space between four spins, i.e the $\SU(2)$-invariant subspace in the tensor product space $\cV^{\otimes 4}$. This is again a two-dimensional space, which we map onto the Hilbert space of an effective qubit:
\be
\dim \textrm{Inv}_{\SU(2)}\,[\cV^{\otimes 4}]
\,=\,2\,.
\ee

We construct a natural basis of this intertwiner space by recoupling the spins  by pair into an internal spin, which can take the value 0 or 1. They are three possible pairing between the four spins, which correspond to three different channels, which we called $s$, $t$, $u$ in the main text and which are illustrated in fig.\ref{mand_intert}. Recoupling the spins 1 and 2 together corresponds to the $s$ channel and gives the following intertwiner basis states:
\begin{align}
\label{intert_state_12}
\ket{0_{s}} &= \frac{1}{2}
	\left( 
		\ket{\uparrow \downarrow \uparrow \downarrow} +
		\ket{\downarrow \uparrow \downarrow \uparrow} - 	
		\ket{\uparrow \downarrow \downarrow \uparrow} -
		\ket{\downarrow \uparrow \uparrow \downarrow}
	\right)	 \nonumber \\
\ket{1_{s}} &= \frac{1}{\sqrt{3}}
		\left(
		\ket{\uparrow \uparrow \downarrow \downarrow } + 
		\ket{\downarrow \downarrow \uparrow \uparrow }
		\right. \nonumber \\
		&-
		\left.
		\frac{
			\ket{\uparrow \downarrow \uparrow \downarrow} + 
			\ket{\downarrow \uparrow \downarrow \uparrow} +
			\ket{\uparrow \downarrow \downarrow  \uparrow} +
			\ket{ \downarrow \uparrow\uparrow \downarrow} 
		}{2}
		\right)
\end{align}
We can similarly express the intertwiner states of spin 0 and 1 in the two other channels $t$ and $u$ in terms of the $\ket{\uparrow,\downarrow}$ states. Then we compute the change of basis between the three channels:
\begin{align}
\begin{pmatrix}
\ket{0_t} \\ \ket{1_t}
\end{pmatrix} &= 
\frac{1}{2}
\begin{pmatrix}
1 & \sqrt{3} \\
 \sqrt{3} & -1
\end{pmatrix}
\begin{pmatrix}
\ket{0_s} \\ \ket{1_s}
\end{pmatrix}
\\
\begin{pmatrix}
\ket{0_u} \\ \ket{1_u}
\end{pmatrix} &= 
\frac{1}{2}
\begin{pmatrix}
-1 & \sqrt{3} \\
- \sqrt{3} & -1
\end{pmatrix}
\begin{pmatrix}
\ket{0_s} \\ \ket{1_s}
\end{pmatrix}
\end{align}

The Casimir operators, given by the scalar product between the $\SU(2)$ generators, are easily calculated in these various basis using the formula $ \vec{J}.\vec{K} = \frac{1}{2}\left[ (\vec{J} + \vec{K})^2 - \vec{J}^2 - \vec{K}^2 \right]$ and the transfer matrices between basis. We compute in the  $(\ket{0}_{12},\ket{1}_{12})$ basis:
\begin{align}
\vec{J_1}.\vec{J_2}
&= 
\frac{1}{4}
\begin{pmatrix}
-3 & 0 \\
0 & 1
\end{pmatrix}
\end{align}
\begin{align}
\vec{J_1}.\vec{J_3}
&= 
\frac{1}{4}
\begin{pmatrix}
0 & -\sqrt{3} \\
 -\sqrt{3} & 2
\end{pmatrix}\,.
\end{align}

This allows to compute the action of the square volume operator $\hat{U}$
as a commutator of those Casimir operators:
\beq
\hat{U}
&=&
-\mi [\vec{J_1}.\vec{J_2}, \vec{J_1}.\vec{J_3}]
\nn\\
&=&
\mi \frac{\sqrt{3}}{4}
\begin{pmatrix}
0 & 1 \\
 -1 & 0
\end{pmatrix}
\eeq

We define the effective qubit states from the eigenvectors of this square volume operator:
\be
\ket{\uud}
\,=\,
\pm \frac{\sqrt{3}}{4}\,
\ket{\uud}\,,
\ee
\begin{align}
\ket{\uud}
	&= \frac{1}{\sqrt{2}} 
	\left( 
		\ket{0_{s}} \mp i \ket{1_{s}}
	 \right)  \\
	 &= \frac{1}{\sqrt{2}} \me^{\mp \frac{\mi \pi}{3}}
	\left( 
		\ket{0_{t}} \pm i \ket{1_{t}}
	 \right)\nn \\
	 &= \frac{1}{\sqrt{2}} \me^{\mp \frac{2 \mi \pi}{3}}
	\left( 
		\ket{0_{u}} \mp i \ket{1_{u}}
	 \right)
		\nn\\
	&= \frac{2}{3} 
	\left( 
		\ket{0_{s}} + \me^{\mp \frac{\mi \pi}{3}} \ket{0_{t}}
		\me^{\mp \frac{\mi 2\pi}{3}} \ket{0_{u}}
	 \right)\nn
\end{align}
These will be the up and down spin states for the Ising model, now corresponding to the two geometrical orientations (of the 3d space $(\R^{3})^{\w 2}\sim\R^{3}$) leading to opposite signs of the squared volume.

In this qubit basis $\ket{\uud}$, we act with the Pauli matrices $\tau^{x,y,z}$ generating the $\su(2)$ algebra. The $\tau^{z}$ operator is naturally defined as the normalized square volume operator:
\be
\tau^{z}\,|\uud\ra
=\pm\,|\uud\ra
=\f{4}{\sqrt{3}}\,\hat{U}\,|\uud\ra\,.
\ee
The $\tau^{x}$ operator is defined from the scalar product operators:
\be
\tau^{x}\,|\uud\ra
=|u_{\downarrow,\uparrow}\ra
=-\left(2\vec{J_1}.\vec{J_2}+\f12
\right)
\,|\uud\ra\,.
\ee
When discussing the low and high temperature definitions of the Ising spin network states, we needed the eigenstates of $\tau^{x}$, which we wrote $\ket{\pm}=(\ket{\uu}\pm\ket{\ud})/\sqrt{2}$. In light of the definition above, these are simply the basis states in the $s$ channel:
\begin{align}
\ket{+} = \ket{0_s} 
\quad
\ket{-} = -\mi\ket{1_s}
\end{align}

\section{Matrix representation of intertwiners}
\label{Matrices_intert}

From the holonomy decomposition of the Ising spin network state, the method
developed to evaluate correlation functions was based on calculating
product of matrices associated to each type of intertwiners, see section 
\ref{loop_representation} for an illustration of the calculation method.
For mathematical completeness, we give  the intertwiner matrices for 
those composed of two (see Figure 
\ref{mand_intert}) or three spin one :

\begin{align}
I_s &=  -\frac{\sqrt{3}}{4}
 	\begin{pmatrix}
		1 & 1/2 & -1/2 \\
		1/2 & 0 & -1/2 \\
		-1/2 & -1/2 & 0
	\end{pmatrix} \nonumber \\
I_t &=  -\frac{\sqrt{3}}{4}
 	\begin{pmatrix}
		0 & 1/2 & 1/2 \\
		1/2 & 1 & 1/2 \\
		1/2 & 1/2 & 0
	\end{pmatrix}  \nonumber \\
I_u &=  -\frac{\sqrt{3}}{4}
 	\begin{pmatrix}
		0 & -1/2 & -1/2 \\
		-1/2 & 0 & 1/2 \\
		-1/2 & 1/2 & 1
	\end{pmatrix} \nonumber \\
I_{111} &=  -\frac{\sqrt{3}}{4} \frac{1}{\sqrt{2}}
 	\begin{pmatrix}
		0 & 1 & 1 \\
		1 & 0 & 1 \\
		1 & 1 & 0
	\end{pmatrix} 
\end{align}
The evaluation of the recoupling of three spin-one is rather simple: there is a unique intertwiner given by the Clebsh-Gordan coefficient. If we see a spin-one as vector, then the invariant subspace of the tensor product of three vectors is given by their scalar triple product (or determinant), which is defined by the   Levi Civita tensor $\epsilon$ (totally antisymmetric tensor).
The intertwiner state  $\ket{I_{111}}$ then reads:
\begin{align}
\ket{I_{111}} = \frac{1}{\sqrt{6}} 
			\sum_{m_i = \pm 1,0}\epsilon^{m_1 m_2 m_3}
			\ket{m_1}\ket{m_2}\ket{m_3}			
\end{align}
where the states $\ket{m}$ of the spin-1 representation are decomposed as tensor products of two spin $1/2$ as:
\begin{align}
\ket{1} &
	   = \ket{\begin{array}{ c }
			\uparrow  \\
			\uparrow 
		   \end{array}}	
\\
\ket{0} &
	   = \frac{1}{\sqrt{2}}
	       \left( 
		\ket{\begin{array}{ c }
			\uparrow  \\
			\downarrow 
			\end{array}}
	  +	\ket{\begin{array}{ c }
			\downarrow  \\
			\uparrow 
			\end{array} }	
	      \right)
\\
\ket{-1} &
	   = \ket{\begin{array}{ c }
			\downarrow  \\
			\downarrow 
		   \end{array}}
\end{align}

Obtaining the matrix is then just a matter of calculationg the different 
scalar products. 
We then have the associated $M$ matrices as defined in the main text:
\begin{align}
\hat{M}_s &= -\frac{3\sqrt{3}}{4} \left(  \frac{1}{2}\ \mathbbm{1}  
			+ \hat{\tau}_x \right)\\
\hat{M}_t &= -\frac{3\sqrt{3}}{4} \left( \frac{1}{2}\ \mathbbm{1}  
			- \cos\left( \frac{\pi}{3}\right) \hat{\tau}_x 
			- \sin\left( \frac{\pi}{3}\right) \hat{\tau}_x  \right) \nonumber \\
\hat{M}_u &= -\frac{3\sqrt{3}}{4} \left( \frac{1}{2}\ \mathbbm{1} 
			+ \cos\left( \frac{2\pi}{3}\right) \hat{\tau}_y 
			+ \sin\left( \frac{2\pi}{3}\right) \hat{\tau}_y \right) \nonumber \\
M_{0111} &= -\frac{\sqrt{3}}{4} 
		\left( 
			\frac{3}{\sqrt{2}} \mathbbm{1} +
			4 \cos\left({\frac{2\pi}{3}}\right)\hat{\sigma}_x -
			4 \sin\left({\frac{2\pi}{3}}\right)\hat{\sigma}_y
		\right) 	\nonumber 
\end{align}

\vfill

\bibliographystyle{bib-style}
\bibliography{IsingStateBiblio}

\end{document}